\documentstyle[12pt,epsfig]{article}
\def\simgt{\rlap{\lower 3.5 pt \hbox{$\mathchar \sim$}} \raise 1pt
  \hbox {$>$}}
\newcommand{\D}{\displaystyle}
\newcommand{\T}{\textstyle}
\begin{document}
\thispagestyle{empty}

\hfill\vbox{\hbox{\bf DESY 94-235}
            \hbox{December 1994}
                                }
\vspace{1.in}
\begin{center}
{\large\bf Gluon Radiation Off Scalar Stop Particles} \\
\vspace{0.5in}
W.~Beenakker, R.~H\"opker and P.~M.~Zerwas \\
\vspace{0.5in}
Deutsches Elektronen-Synchrotron DESY, D-22603 Hamburg, FRG \\
\end{center}
\vspace{2in}

\begin{center}
ABSTRACT \\
\end{center}
We present the distributions for gluon radiation off stop-antistop
particles produced in $e^+e^-$ annihilation: $e^+e^- \to
\tilde t\, \bar{\tilde t} \,g$.  For high energies the splitting
functions of the
fragmentation processes $\tilde t \to \tilde t \,g$ and $g \to \tilde t
\,\bar{\tilde t}$ are derived; they
are universal and apply also to high-energy stop particles produced at
hadron colliders.

\pagebreak

\paragraph{Introduction.}
Stop particles are exceptional among the supersymmetric partners of
the standard-model fermions. Since the top quarks are heavy, the
masses of the two stop particles $\tilde t_1$ and $\tilde t_2$,
mixtures of the left (L) and right (R)
squarks, may split into two levels separated by a large gap
\cite{ellis}-\cite{olive}. The mass of the lightest
eigenstate $\tilde t_1$
could be so low that the particle may eventually be accessible at
the existing $p\bar p$ and even $e^+e^-$ storage rings. So far the result
of search experiments at $e^+e^-$ colliders \cite{tristan,opal} has
been negative and a lower limit of $45.1$ GeV has been set at
LEP \cite{opal} for the L/R mixing angle outside the band of
$\cos^2\theta_t$ between 0.17 and 0.44
and for a mass difference between the
$\tilde t_1$ and the lightest neutralino $\tilde\chi_1^0$ of more
than $5$ GeV. The higher energy at LEP2 and
dedicated efforts at the Tevatron will open the mass range beyond the
current limits soon.

\begin{figure}[b]
  \begin{center}
    \leavevmode
    \epsfig{file=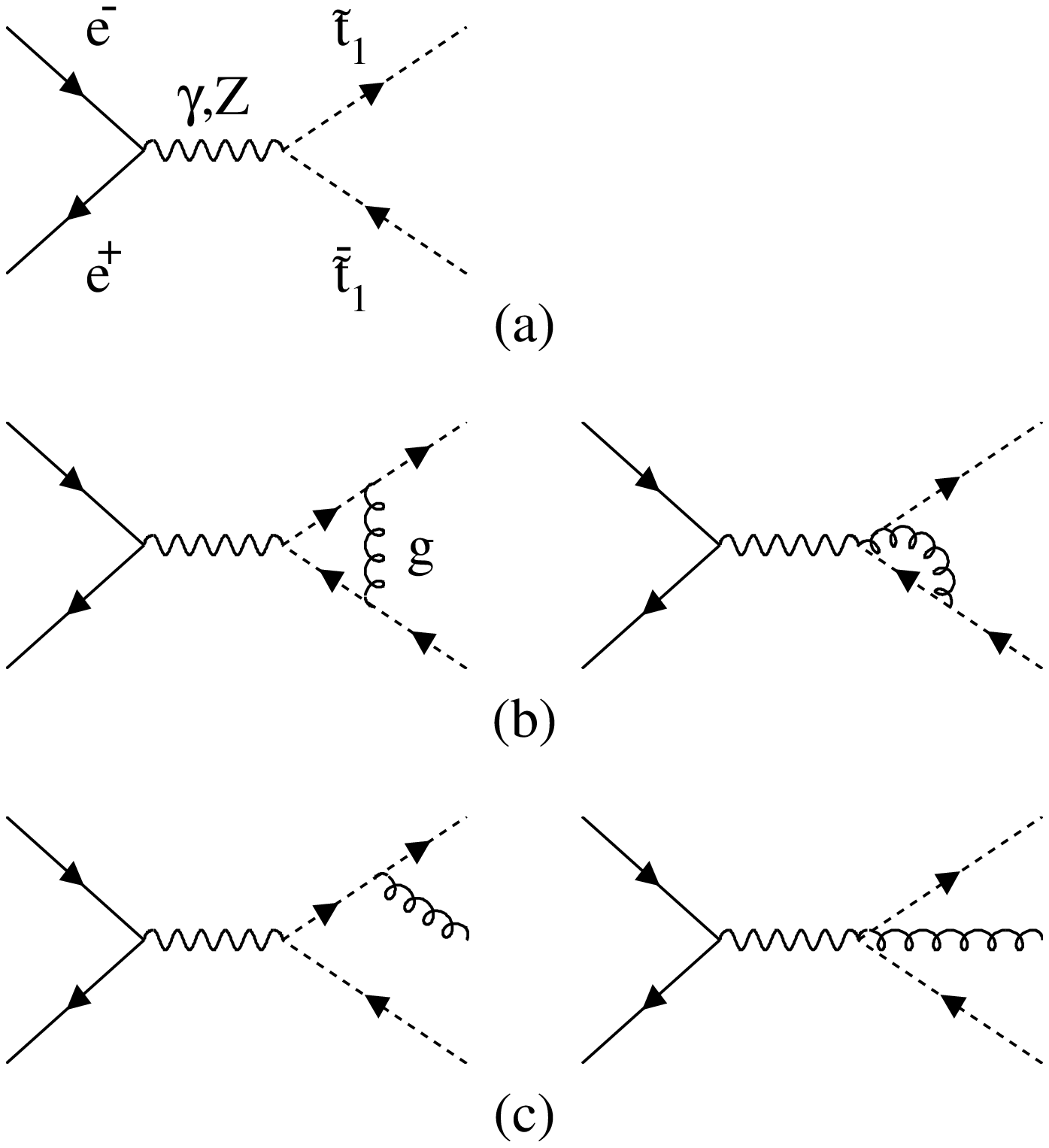,height=12cm,bbllx=3.1cm,bblly=6cm,%
bburx=18cm,bbury=18cm}
  \end{center}
  \caption{Generic diagrams for $\tilde t_1 \bar{\tilde t}_1$
    production in $e^+e^-$ collisions. (a) Born level; (b) virtual
    QCD corrections; (c) gluon radiation.}
  \label{feyn}
\end{figure}

To begin, we briefly summarize the well-known theoretical predictions for
the cross section  of the production process [Fig.\ref{feyn}(a)]
\begin{displaymath}
  e^+\,e^- \to \tilde t_1\, \bar{\tilde t}_1
\end{displaymath}
For a given value $\theta_t$ of
the L/R mixing angle, the vertices of the
$\tilde t_1$ pair with the photon and the $Z$ boson may be written as
$i e_0 \tilde Q [p_{\tilde t_1} - p_{\bar{\tilde t}_1}]_\mu$,
where $p_{\tilde t_1}$ and $p_{\bar{\tilde t}_1}$ are the 4-momenta of
the stop and antistop squarks, and the charges read
\begin{eqnarray*}
  \tilde Q_\gamma & = & - e_t \\
  \tilde Q_Z & = & (\cos^2\theta_t -2 e_t \sin^2\theta_W)/\sin 2\theta_W
\end{eqnarray*}
respectively. $\theta_W$ is the standard electroweak mixing
angle and $e_0=\sqrt{4\pi\alpha}$ is the electromagnetic coupling to be
evaluated with $\alpha^{-1}(M_Z) = 129.1$ in the
improved Born approximation \cite{swartz}. The
$Z$ boson coupling vanishes for the L/R mixing angle
$\cos^2 \theta_t \to 2 e_t \sin^2 \theta_W \approx 0.30$. Defining the
$\gamma$ and $Z$ vector/axial-vector charges of the electron, as
usual, by $e_e = -1$, $v_e = -1 +4 \sin^2 \theta_W$ and $a_e =
-1$, the cross section can be expressed in the compact form \cite{hikasa}
\begin{eqnarray}
  \sigma_B[e^+e^- \to \tilde t_1 \bar{\tilde t}_1] &=&
  \frac{\pi\alpha^2}{s} \left[\tilde Q_\gamma^2 + \frac{(v_e^2
    +a_e^2)\tilde Q_Z^2}{4\sin^2 2\theta_W}
  \frac{s^2}{(s -M_Z^2)^2 + M_Z^2 \Gamma_Z^2}\right. \nonumber\\
  & &\hphantom{\frac{\pi\alpha^2\beta^3}{s}a}\left.+\,\frac{v_e\tilde
    Q_\gamma \tilde Q_Z}{\sin2\theta_W} \frac{s (s-M_Z^2)}{(s
    -M_Z^2)^2 + M_Z^2 \Gamma_Z^2}\right] \beta^3
\label{born}
\end{eqnarray}
where $\sqrt{s}$ is the center of mass
energy and $M_Z,\,\Gamma_Z$ are the mass and the total
width of the $Z$ boson, respectively.
The $P$-wave excitation near the threshold gives rise to the familiar
$\beta^3$ suppression, where $\beta = (1 -4 m^2_{\tilde t_1}/s)^{1/2}$ is
the velocity of the stop particles. Angular momentum conservation
enforces the $\sin^2\theta$ law, $\sigma_B^{-1} d\sigma_B/d\cos\theta
= \frac{3}{4}\sin^2\theta$, for the angular distribution of the stop
particles with respect to the beam axis.

\paragraph{QCD corrections.}
Gluonic corrections modify the cross section \cite{schwinger,
drees}\footnote{Since we focus on QCD gluon effects for light stop
          particles
          in the LEP range, we do not take into account quark-gluino
          loop effects, assuming the gluino to be heavy; these
          loop effects have been discussed for squark production at the
          Tevatron in Ref.\cite{been} and at $e^+e^-$ colliders in
          Ref.\cite{djou}.}.
The virtual
corrections, Fig.\ref{feyn}(b), can be expressed by the form factor
\newpage
\begin{eqnarray}
  F(s) &=& \frac{4}{3}\frac{\alpha_s}{\pi}\left\{\frac{s -2 m_{\tilde
      t_1}^2}{s\beta}\left[2\, \mbox{Li}_2(w)
  + 2\log(w)\log(1-w) - \frac{1}{2}\log^2(w)
  +\frac{2}{3}\pi^2 - 2\log(w) \right.\right.\nonumber \\
  & & \hphantom{\frac{4}{3}\frac{\alpha_s}{\pi}a}\left.\left.
  -\,\log(w)\log\left(\frac{\lambda^2}{m_{\tilde t_1}^2}\right)\right]
   -2 - \log\left(\frac{\lambda^2}{m_{\tilde t_1}^2}\right)\right\}
\end{eqnarray}
where $\alpha_s$ is the strong coupling constant and  the
kinematical variable $w$ is defined as $w =(1-\beta)/(1+\beta)$.
The form factor is infrared (IR) divergent. We have
regularized this divergence by introducing a small parameter
$\lambda$ for the gluon mass. The IR
singularity is eliminated by adding the contribution of the soft gluon
radiation [Fig.\ref{feyn}(c)], with the scaled gluon energy
integrated up to a cut-off value $\epsilon_g = 2
E_g^{cut}/\sqrt{s} \ll 1$. The sum of the virtual correction
($V$) and the soft-gluon radiation ($S$) depends
only on the physical energy cut-off $\epsilon_g$,
\begin{eqnarray*}
  \sigma_{V+S} & = & \sigma_B\frac{4}{3}\frac{\alpha_s}{\pi}\left\{\frac{s -2
      m_{\tilde t_1}^2}{s\beta}\left[\vphantom{\frac{1}{3}}
    4\,\mbox{Li}_2(w)  -2\log(w)\log(1+w)
     + 4\log(w)\log(1-w) \right. \right. \\
    & &\left. \left.  + \,\frac{1}{3}\pi^2
    -2\log(w)\log(\epsilon_g) \right]  +
    \frac{4m_{\tilde t_1}^2 - 3 s}{s \beta}\log(w) +
 \log\left(\frac{m_{\tilde t_1}^2}{s}\right) -2\log(\epsilon_g) -2 \right\}
\end{eqnarray*}

\begin{figure}[b]
  \begin{center}
    \leavevmode
    \epsfig{file=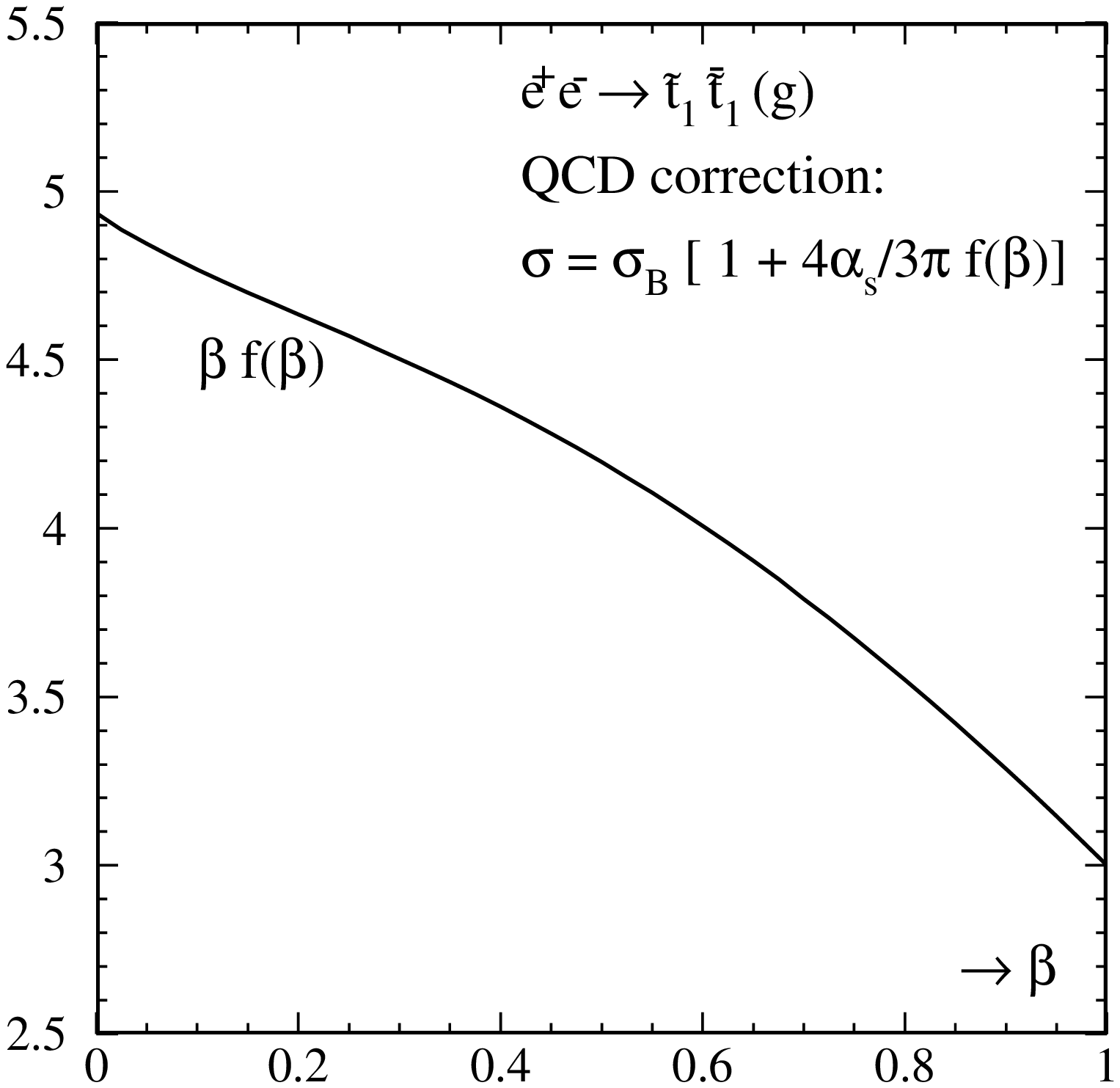,height=12cm,bbllx=3.1cm,bblly=5cm,%
bburx=18cm,bbury=18cm}
   \end{center}
  \caption{Coefficient of the QCD correction to the total cross
    section; shown is $\beta f(\beta)$, cf. eq.(\protect\ref{sigma}),
    with $\beta = (1 -4m_{\tilde t_1}^2/s)^{1/2}$.}
  \label{fbeta}
\end{figure}

After including the hard gluon radiation, the
dependence on the cut-off $\epsilon_g$ disappears from the total cross
section. The total QCD corrections
can finally be summarized in a universal
factor \cite{drees}
\begin{equation}
  \sigma[  e^+e^- \to \tilde t_1 \,\bar{\tilde t}_1\, (g)]
 = \sigma_B \left[1 +\frac{4}{3}\frac{\alpha_s}{\pi} f(\beta)\right]
\label{sigma}
\end{equation}
with (Fig.\ref{fbeta})
\begin{eqnarray*}
  f(\beta) &=& \frac{1+\beta^2}{\beta}\left[ \vphantom{\frac{1}{3}}
  4\,\mbox{Li}_2(w) +
  2\,\mbox{Li}_2(-w) +2\log(w)\log(1-w) +\log(w)\log(1+w)\right] \nonumber\\
 & & -\,4\log(1-w) -2\log(1+w) +\left[3 + \frac{1}{\beta^3}\left(2
 -\frac{5}{4}(1 + \beta^2)^2\right)\right]\log(w)
 +\frac{3}{2}\frac{1+\beta^2}{\beta^2}
\end{eqnarray*}
Very close to the threshold the Coulombic gluon exchange between the
slowly moving stop particles generates the universal Sommerfeld
rescattering singularity \cite{sommer} $f \to \pi^2/2\beta$, which damps the
threshold suppression, yet does not neutralize it entirely.
Employing methods based on non-relativistic Green's functions,
an adequate
description of stop pair production near threshold has been given in
Ref.\cite{bigi1}, which also
takes into account screening effects due to the finite decay width of
the stop particles. In the high-energy limit \cite{drees} the
correction factor in eq.(\ref{sigma}) approaches the value $( 1 +
4\alpha_s/\pi)$.

In this note we present a general analysis of hard gluon radiation. We
also include stop fragmentation due to collinear gluon emission in the
perturbative regime at high energies and we give
an account of non-perturbative fragmentation effects.

\newpage
For unpolarized lepton beams the cross section for gluon radiation
off $\tilde t_1$ squarks
\begin{displaymath}
  e^+\,e^- \to \tilde t_1\, \bar{\tilde t}_1 \,g
\end{displaymath}
depends on four variables: the polar angle $\theta$ between the
momentum of the $\tilde t_1$ squark and the $e^-$ momentum, the
azimuthal angle $\chi$ between the $\tilde t_1 \bar{\tilde t}_1 g $ plane
and the plane spanned by the $e^\pm$ beam axis with the $\tilde t_1$
momentum [see Ref.\cite{laer1}], and two of the scaled
energies $x(\tilde t_1)$, $\bar{x}(\bar{\tilde t_1})$, $z(g)$ in units of
the beam energy. The energies are related through $x +\bar x +z
=2$ and vary
over the intervals $\mu \leq x,\bar x \leq 1$ and $0\leq z \leq 1 -\mu^2$,
where $\mu = 2m_{\tilde t_1}/\sqrt{s}$ denotes the squark
mass in units of the beam energy. For the angles between the squark
and gluon momenta we have
\begin{eqnarray*}
  \cos\theta_{\tilde t_1 \bar{\tilde t}_1} & = & \frac{2 -2(x +\bar x)
    + x \bar
  x +\mu^2}{\sqrt{(x^2-\mu^2)(\bar x^2 -\mu^2)}}\\
  \cos\theta_{\tilde t_1 g} & = &\frac{2 -2 (x +z) +x
    z}{z\sqrt{x^2-\mu^2}}
\end{eqnarray*}
The spin-1 helicity analysis of the cross
section results in the following well-known angular decomposition \cite{laer2}
\begin{eqnarray}
  \frac{d\sigma}{dx\,d\bar x\,d\cos\theta\,d\chi/2\pi} &=&
  \frac{3}{8}( 1+ \cos^2\theta) \frac{d\sigma^U}{dx\,d\bar x}
  +\frac{3}{4}\sin^2\theta \frac{d\sigma^L}{dx\,d\bar x} \nonumber\\
  & & -\frac{3}{2\sqrt{2}}\sin2\theta\cos\chi
  \frac{d\sigma^I}{dx\,d\bar x}
  +\frac{3}{4}\sin^2\theta \cos 2\chi \frac{d\sigma^T}{dx\,d\bar x}
\end{eqnarray}
[$U$ = transverse (no flip), $L$ = longitudinal, $I$ = trv$\ast$long,
$T$ = trv$\ast$trv (flip)].
If the polar and azimuthal angles are integrated out, the cross
section is given by $\sigma = \sigma^U +\sigma^L$.

It is convenient to write the helicity cross sections as
\begin{equation}
  \frac{\beta^3}
       {\sigma_B}\frac{d\sigma^j}{dx\,d\bar x} = \frac{\alpha_s}{4\pi}
       \frac{S^j + \mu^2 N^j}{(1-x)(1-\bar x)}
\label{d2sigma}
\end{equation}
The densities $S^j$ and $N^j$ are summarized in Table \ref{sandn};
$p$ is the momentum of the $\tilde t_1$ squark, $\bar p$
and $k$ are the longitudinal momenta of
$\bar{\tilde t}_1$ and $g$ in the $\tilde t_1$ direction, and $p_T$
is the modulus of the transverse $\bar{\tilde t}_1$, $g$ momentum with
respect to this axis [all momenta in units of the beam energy].
Since $I,T$ correspond to $\gamma,Z$ helicity flips by 1
and 2 units, they are of order $p_T$ and $p_T^2$, respectively.
Note that the threshold suppression is absent in the $U$, $I$, $T$
components and attenuated in the leading longitudinal $L$ term as
expected from eq.(\ref{sigma}).

\begin{table}[t]
  \begin{center}
    \leavevmode
    \begin{tabular}{|c|c|c|}
      \hline
      & & \\[-0.2cm]
      & $S^j$ & $N^j$ \\
      & & \\[-0.2cm]
      \hline
      & & \\[-0.2cm]
      $\D  U $ & $\D \frac{\T32}{\T3}\,(1-x)\,(1-\bar x)$  &
      $\D -\frac{\T4}{\T3}\,p_T^2\,\frac{\T1-x}{\T1-\bar x}$ \\ [0.4cm]
      $\D  L $ & $\D \frac{\T16 \beta^2}{\T3}\,(1-z) $ &
      $\D \frac{\T4}{\T3}\,\left[p_T^2\,
      \frac{\T1-x}{\T1-\bar x} - \beta^2 \left(\frac{\T1-x}{\T1-\bar x}
      +\frac{\T1-\bar x}{\T1-x} +2\right)\right] $ \\ [0.4cm]
      $\D  I $ & $\D  -\frac{\T4\sqrt{\T2}}{\T3}\,p_T \,p$ &
      $\D \frac{\T2\sqrt{\T2}}{\T3}\,p_T \left(p - \bar p\,
      \frac{\T1-x}{\T1-\bar x}\right) $\\ [0.4cm]
      $\D  T $ & $\D 0$& $\D \frac{\T2}{\T3}\,p_T^2\,
      \frac{\T1-x}{\T1-\bar x} $ \\[-0.2cm]
      & & \\
      \hline
    \end{tabular}
  \end{center}
  \caption{Coefficients of the helicity cross sections in
    eq.(\protect\ref{d2sigma}). The energy and momentum variables are
    defined in the text.}
  \label{sandn}
\end{table}

\paragraph{Fragmentation.}
In the limit where the gluons are emitted from fast moving squarks
with small angles, the gluon radiation
\begin{displaymath}
  \label{brems}
  \tilde t_1 \to \tilde t_1\, g
\end{displaymath}
can be interpreted as a perturbative fragmentation process. From
the form of the differential cross section $d\sigma/dz\,dp_T^2 $ we
find in this limit for the splitting functions, in
analogy to the Weizs\"acker-Williams \cite{weiz} and Altarelli-Parisi
splitting functions \cite{altarelli},
\begin{eqnarray}
  P[\tilde t_1 \to \tilde t_1; x] & = & \frac{\alpha_s}{2\pi}\,
  \frac{8}{3}\, \frac{x}{1-x}\, \log\frac{Q^2}{m_{\tilde t_1}^2}  \\
  P[\tilde t_1 \to g; z] & = & \frac{\alpha_s}{2\pi}\, \frac{8}{3}\, \frac{
    1-z}{z}\, \log\frac{Q^2}{m_{\tilde t_1}^2}
                                                                 \nonumber
\end{eqnarray}
As usual, $x$ and $z$ are the fractions of energy
transferred from the $\tilde t_1$ beam to the squark $\tilde t_1$ and
the gluon $g$ after
fragmentation, respectively; $Q$ is the
evolution scale of the elementary process, normalized by the squark
mass rather than the QCD $\Lambda$ parameter [in contrast to the light
quark/gluon sector]. As a consequence of angular-momentum
conservation, the gluon cannot pick up the total momentum of
the squark beam. [Similar zeros have been found for helicity-flip
fragmentation functions in QED/ QCD \cite{altarelli,falk}.]

By using the crossing
rules \{$z \to 1, 1 \to x \}\mbox{ and }\{ 1-x \leftrightarrow 1-x$\},
familiar from the analogous splitting functions in QED \cite{chen}, we
derive for the elementary gluon splitting process into a
squark-antisquark pair
\begin{displaymath}
g \to \tilde t_1 \bar{\tilde t_1}
\end{displaymath}
the distribution
\begin{equation}
  P[g \to \tilde t_1; x] =  \frac{\alpha_s}{2\pi}\, \frac{1}{2} \,x \,(1-x)
  \,\log\frac{Q^2}{m_{\tilde t_1}^2}
\end{equation}
after adjusting color and spin coefficients properly. This splitting
function is symmetric under the $\tilde t_1 \leftrightarrow
\bar{\tilde t_1}$ exchange, i.e. $\{x \leftrightarrow
1-x\}$. The probability is maximal for the splitting into equal
fractions $x = 1/2$ of the momenta, in contrast to spinor QED/QCD
where the splitting into a quark-antiquark pair is proportional
to $x^2 + (1-x)^2$ and hence asymmetric configurations are preferred.

The above splitting functions provide the kernels for the
shower expansions in
perturbative QCD Monte Carlos for $e^+e^-$ annihilation such as Pythia
\cite{sjos} and Herwig \cite{webb}. They serve the same purpose in the
hadron-hadron versions of these generators as well as Isajet
\cite{baer}. Of course, the interpretation of the
radiation processes as universal fragmentation processes becomes
increasingly
adequate with rising energy of the fragmenting squarks/gluons.

If the $\tilde t_1$ squark is lighter than the top quark, the
lifetime will be long, $\tau \ge 10^{-20}$sec, since the dominant decay
channel $\tilde t_1  \to t +\tilde\chi_1^0$ is shut off
$[\tilde\chi_1^0 = LSP]$. The decay widths corresponding to
the 2-body decay $\tilde t_1 \to c +\tilde\chi_1^0$ and 3-body
slepton decays involve the
electroweak coupling twice and hence will be very small
\cite{hikasa}. As a result, the lifetime is much
longer than the typical non-perturbative fragmentation time of order
$1$ fm [i.e. ${\cal O}\, (10^{-23}$sec)] so that the squark has got
enough time to form $(\tilde t_1\bar q)$
and $(\tilde t_1 q q)$ fermionic and bosonic hadrons. However, the
energy transfer due to the non-perturbative
fragmentation, evolving after the early perturbative
fragmentation, is very small as a result of Galilei's law of
inertia. Describing this last step in the hadronization process of a
$\tilde t_1$ jet by the non-perturbative fragmentation function {\it
  \`a la} Peterson et al. \cite{pete}
(which accounts very well for the heavy-quark analogue), we find
\begin{equation}
  D (x)^{NP} \approx \frac{4 \sqrt{\epsilon}}{\pi}\,\frac{1}{x\,[1 - 1/x
    -\epsilon/(1-x)]^2}
\end{equation}
with the parameter $\epsilon \sim 0.5\,\mbox{GeV}^2/m_{\tilde
  t_1}^2$. Here, $x = E[(\tilde t_1\,\bar q)]/E[\tilde t_1]$ is the
energy fraction transferred from the $\tilde t_1$ parton to the $(\tilde
t_1\,\bar q)$ hadron etc. The resulting average non-perturbative energy loss
\begin{displaymath}
<1-x>^{NP}\sim\frac{2\sqrt{\epsilon}}{\pi}\left[
\log\left(\frac{1}{\epsilon}\right) -3\right]
\end{displaymath}
is numerically at the level of a few percent.

Monte Carlo programs for the hadronization of $\tilde t_1$ squarks
link the early perturbative fragmentation with the subsequent
non-perturbative hadronization. The relative weight of perturbative
and non-perturbative fragmentation can be characterized by the
average energy loss in the two consecutive steps. The overall retained
average energy
of the $\tilde t_1$ squarks factorizes into the two components,
\begin{equation}
<x> = <x>^{NP} <x>^{PT}
\end{equation}
Summing up the energy loss due to multiple gluon radiation
at high energies, we
find in analogy to heavy-quark fragmentation \cite{bigi2}
\begin{displaymath}
  <x>^{PT} = \left[\frac{\alpha_s(m_{\tilde
      t_1}^2)}{\alpha_s(E^2)}\right]^{-8/3b}
\end{displaymath}
with $ b = (11 -2 n_f/3) +(-2 - n_f/3)$ being the LO QCD $\beta$
function including the colored supersymmetric particle spectrum.
At high energies, the perturbative multi-gluon radiation has a
bigger impact than the final non-perturbative
hadronization mechanism, e.g. $<x>^{PT} \approx 0.93$ for a $\tilde
t_1$ beam energy $E=1\,\mbox{TeV}$ and $m_{\tilde t_1}=200\,\mbox{GeV}$
as compared to $<x>^{NP}\approx 0.98$. At low energies the two
fragmentation effects are of comparable size.

After finalizing the manuscript, we received a copy of Ref.\cite{djou}
in which the total cross sections for squark pair production in $e^+e^-$
annihilation have been discussed including squark-gluon and quark-gluino
loops, yet not the gluon-jet distributions analysed in the present note.

We thank our colleagues at the LEP2 Workshop who demanded the analysis
presented here to refine the experimental stop search techniques.

\pagebreak

\pagebreak


\begin{thebibliography}{xx}
\bibitem{ellis} J. Ellis and S. Rudaz, Phys. Lett. {\bf B128}
(1983) 248.
\bibitem{hikasa} K. Hikasa and M. Kobayashi, Phys. Rev. {\bf D36}
(1987) 724.
\bibitem{olive} K.A. Olive and S. Rudaz, Phys. Lett. {\bf B340} (1990)
                74.
\bibitem{tristan} J. Shirai {\it et al.} (Venus), Phys. Rev.
Lett. {\bf 72} (1994) 3313.
\bibitem{opal} R. Akers {\it et al.} (Opal), Phys. Lett. {\bf
B337} (1994) 207.
\bibitem{swartz} J.M.L. Swartz, SLAC-PUB-6710 (Nov. 1994).
\bibitem{schwinger} J. Schwinger, {\it Particles, Sources and Fields}
                    vol.~II (Addison-Wesley, New York 1973).
\bibitem{drees} M. Drees and K. Hikasa, Phys. Lett. {\bf B252} (1990)
127.
\bibitem{been} W. Beenakker, R. H\"opker, M. Spira and P.M. Zerwas,
               Report DESY 94-212.
\bibitem{djou} A. Arhrib, M. Capdequi-Peyranere and A. Djouadi, Montreal
               Report UdeM-GPP-94-13.
\bibitem{sommer} A. Sommerfeld, {\it Atombau und Spektrallinien} vol.~2
  (Vieweg, Braunschweig 1939).
\bibitem{bigi1} I.I. Bigi, V.S. Fadin and V. Khoze, Nucl. Phys. {\bf
    B377} (1992) 461.
\bibitem{laer1} E. Laermann, K.H. Streng and P.M. Zerwas, Z. Phys.
{\bf C3} (1980) 289.
\bibitem{laer2} E. Laermann and P.M. Zerwas, Phys. Lett. {\bf B89}
(1980) 225.
\bibitem{weiz} C.F.v. Weizs\"acker, Z. Phys. {\bf 88} (1934) 612;
E.J. Williams, Phys. Rev. {\bf 45} (1934) 729.
\bibitem{altarelli} G. Altarelli and G.
Parisi, Nucl. Phys. {\bf B126} (1977) 298.
\bibitem{falk} B. Falk and L.M. Sehgal, Phys. Lett. {\bf B325}
(1994) 509.
\bibitem{chen} M.-S. Chen and P.M. Zerwas, Phys. Rev. {\bf D12}
(1975) 187.
\bibitem{sjos} T. Sj\"ostrand, Comp. Phys. Commun. {\bf 82} (1994) 74.
\bibitem{webb} G. Marchesini {\it et al.}, Comp. Phys.
Commun. {\bf 67} (1992) 465; B.R. Webber, Cavendish-HEP-94/17.
\bibitem{baer} H. Baer, F. E. Paige, S. D. Protopopescu and X. Tata,
  Report on ISAJET 7.0/ ISASUSY 1.0, FSU-HEP 930329 and UH-511-764-93.
\bibitem{pete} C. Peterson, D. Schlatter, I. Schmitt and P.M. Zerwas,
Phys. Rev. {\bf D27} (1983) 105.
\bibitem{bigi2} I.I. Bigi, Yu.L. Dokshitser, V. Khoze, J.H. K\"uhn and
P.M. Zerwas, Phys. Lett. {\bf B181} (1986) 157.
\end{thebibliography}
\end{document}